\documentclass[prd,twocolumn,showpacs, letter]{revtex4}
\usepackage{epsfig}
\usepackage{graphicx}
\usepackage{amsmath}
\usepackage{color}
\usepackage{fancyvrb}
\usepackage{longtable}
\newcommand{\Hethree}{$^3{\rm {He}}$}

\newcommand{\MHz}{{\rm {MHz}}}
\newcommand{\GHz}{{\rm {GHz}}}

\newcommand{\Hethreeplus}{^3{\rm He}^+}

\newcommand{\Hethreesplitting}{$8.7$~GHz}
\newcommand{\Hethreesplittingprec}{$8.666$~GHz}
\newcommand{\Hethreeterm}{$^2 {\rm S}_{1/2}$ $F=0-1$}
\newcommand{\Tcmb}{T_{\rm CMB}}
\newcommand{\Tspin}{T_{\rm s}}
\newcommand{\Tkinetic}{T_{\rm k}}
\newcommand{\kB}{k_{\rm B}}

\newcommand{\xHeII}{x_{\rm HeII}}
\newcommand{\xHeIII}{x_{\rm HeIII}}
\newcommand{\nH}{n_{\rm H}}
\newcommand{\nelectron}{n_{\rm e}}

\newcommand{\HI}{H{\sc ~i}}
\newcommand{\HII}{H{\sc ~ii}}
\newcommand{\HeI}{He{\sc ~i}}
\newcommand{\HeII}{He{\sc ~ii}}
\newcommand{\HeIII}{He{\sc ~iii}}

\newcommand{\red}[1]{}

\newcommand{\comment}[1]{}
\newcommand{\beq}{\begin{equation}}
\newcommand{\eeq}{\end{equation}}
\newcommand{\beqa}{\begin{eqnarray}}
\newcommand{\eeqa}{\end{eqnarray}}
\newcommand{\barr}{\begin{array}}
\newcommand{\earr}{\end{array}}

\begin{document}
\title{Redshifted intergalactic $\Hethreeplus$ \Hethreesplitting\ hyperfine absorption}

\author{Matthew McQuinn}
\email{mmcquinn@cfa.harvard.edu}
\affiliation{Harvard-Smithsonian Center for Astrophysics, 60 Garden St., Cambridge, MA 02138, USA}
\author{Eric R. Switzer}
\email{switzer@kicp.uchicago.edu}
\affiliation{Kavli Institute for Cosmological Physics, The University of Chicago, Chicago, IL, 60637, USA}
\date{\today}

\begin{abstract}
Motivated by recent interest in redshifted $21\,$cm emission of intergalactic hydrogen, we investigate the \Hethreesplitting\ \Hethreeterm\ hyperfine transition of $\Hethreeplus$.  While the primordial abundance of $^3{\rm He}$ relative to hydrogen is $10^{-5}$, the hyperfine spontaneous decay rate is $680$ times larger.  Furthermore, the antenna temperature is much lower at the frequencies relevant for the $\Hethreeplus$ transition compared to that of $z> 6$ $21\,$cm emission.  We find that the spin temperature of this $8.7\;$GHz line in the intergalactic medium is approximately the cosmic microwave background temperature, such that this transition is best observed in absorption against high-redshift, radio-bright quasars.   We show that intergalactic $8.7\;$GHz absorption is a promising, unsaturated observable of the ionization history of intergalactic helium (for which \HeII$\rightarrow$\HeIII\ reionization is believed to complete at $z \sim 3$) and of the primordial $^3{\rm He}$ abundance.  Instruments must reach $\sim 1 \, \mu$Jy RMS noise in bands of $1\,$MHz on a $1\,$Jy source to directly resolve this absorption.  However, in combination with \HI\ Ly$\alpha$ forest measurements, an instrument can statistically detect this absorption from $z>3$ with $30 \,\mu$Jy RMS noise in $0.1\,$MHz spectral bands over $100$~MHz, which may be within the reach of present instruments.
\end{abstract}
\pacs{95.30.Dr, 32.10.Fn, 98.62.Ra}
% 98.62.Ra Intergalactic matter; quasar absorption and emission-line systems
% 95.30.Dr Atomic processes and interactions 
% 32.10.Fn Fine and hyperfine structure 

\maketitle

\section{Introduction}
\label{sec:intro}

The $^2 {\rm S}_{1/2}$ $F=0-1$ hyperfine splitting of neutral hydrogen at $1.42$~GHz has been the subject of intense study as an observable of the cosmological reionization of hydrogen.  Several experiments aim to map the high-redshift $1.42$~GHz emission, but another approach is to study the pre-reionization era in absorption against bright radio sources \cite{2002ApJ...577...22C, 2002ApJ...579....1F, 2006MNRAS.370.1867F}.  This approach is restricted to times when the spin temperature of intergalactic hydrogen is close to the cosmic microwave background (CMB) temperature ($z \gtrsim15$) and may prove difficult because of the paucity of bright radio sources from this era.
 
Here, we propose a similar experiment using the $\Hethreeplus$ \Hethreeterm\ hyperfine transition at \Hethreesplittingprec\ \cite{1969PhRv..187....5S, 1967ApJ...149...15G} in absorption.  This proposal is motivated by the fact that, while the abundance of $^3{\rm He}$ is $\sim 10^{-5}$ that of hydrogen, the spontaneous decay rate $A_{10}$ is $680$ times larger ($A_{10} =1.95 \times 10^{-12}~{\rm s}^{-1}$ for $\Hethreeplus$ \cite{1994ApJ...423..522G}) and because we find the spin temperature of this transition is close to that of the CMB, making $z\sim3$ intergalactic $8.7\;$GHz emission very difficult to observe.  We show that intergalactic $8.7\;$GHz absorption can be used as an observable of cosmological \HeII$\rightarrow$\HeIII\ reionization (hereafter, ``\HeII\ reionization") and \HeI$\rightarrow$\HeII\ reionization (which is believed to be concurrent with \HI\ reionization).   This absorption can also provide a handle on the abundance of $^3{\rm He}$ from primordial nucleosynthesis.

The reionization of the hydrogen and helium in the intergalactic medium is one of the least understood events in our cosmic history. Observations of the Ly$\alpha$ forest constrain the intergalactic hydrogen to be reionized at $z>6$.  For any plausible astrophysical spectrum, if it ionizes the hydrogen, then helium also becomes at least singly ionized.  However, it takes a hard source of ultraviolet radiation to doubly ionize the helium.  Quasars are an intense source of such radiation, and their abundance peaks at $z\sim 3$.  The present paradigm is that they finish doubly ionizing the helium at $z\sim 3$  \cite{2008ApJ...681....1F, 2008ApJ...682...14F, 2008ApJ...686...25F, 2009ApJ...694..842M}.  However, helium becomes doubly ionized at the same time that hydrogen is reionized (at $z>6$) in more exotic models, such as if dark matter annihilation byproducts \cite{2009arXiv0904.1210B} or black holes  \cite{2009arXiv0905.0144V} ionize the hydrogen.

Several observations of the IGM suggest that \HeII\ reionization is ending at $z\sim 3$. \citet{2000ApJ...534...41R} and \citet{2000MNRAS.318..817S} measured the gas temperature from the widths of the narrowest lines in the \HI\ Ly$\alpha$ forest and found an increase in the temperature of $\Delta T \sim 10^4$~K between $z = 3.5$ and $z = 3$ (before a decline at lower redshifts) that is likely due to the heating from \HeII\ reionization.   In addition, observations of \HeII\ Ly$\alpha$ absorption at $2.8 < z < 3.3$ find $10$s of comoving Mpc (cMpc) regions with no detected transmission \cite{heap00}, which may signify the presence of diffuse intergalactic \HeII\ \cite{2009arXiv0905.0481M}.  Finally, \citet{1998AJ....115.2184S} and \citet{2005A&A...441....9A} interpreted evolution in the column density ratios of certain highly ionized metals at $z \approx 3$ as evidence for \HeII\ reionization.  However, all of these probes are quite indirect and their interpretation is controversial.   The observable discussed in this paper, $\Hethreeplus$ hyperfine absorption, would be a more direct probe of \HeII\ reionization.

Observations of $\Hethreeplus$ hyperfine emission from \HII\ regions are currently the most reliable method to constrain primordial and stellar production of $^3{\rm He}$ \cite{1967ApJ...149...15G, 1995ApJ...444..680O,1998SSRv...84..185R}.  Such observations have constrained the fractional abundance of $^3{\rm He}$ to be $1.5^{+1.0}_{-0.5} \times 10^{-5}$ \cite{1998SSRv...84..185R}, consistent with the Big Bang Nucleosynthesis value of $^3{\rm He/H} = 1.04 \pm .04 \times 10^{-5}$ implied by the best-fit WMAP cosmology \citep{2009ApJS..180..330K}.  These galactic $\Hethreeplus$ measurements require detailed modeling of \HII\ regions and their emission lines.  In addition, traditional models predict that stellar nucleosynthesis should produce a significant quantity of $^3{\rm He}$ in excess of the primordial abundance.  Surprisingly, in most \HII\ regions, a significant excess is not found (e.g. \cite{2008ApJ...677..581E, 2007Sci...317.1171B}).  An intergalactic measurement of the $^3{\rm He}$ would be much less contaminated by $^3{\rm He}$ produced in stars, and such a measurement is likely required for a precision measurement of the primordial $^3{\rm He}$ abundance.

Previous studies have mentioned $\Hethreeplus$ hyperfine transition as a potential cosmological observable \cite{1966AZh....43.1237S, 2006PhRvL..97i1301S, 2006PhR...433..181F, 2007AstL...33...67S, 1984SvAL...10..201S}.  \citet{1966AZh....43.1237S} pointed out that this line exists and could potentially be used to constrain big bang nucleosynthesis.  \citet{2006PhRvL..97i1301S} noted that it might be easier to observe than the $92\;$cm hyperfine line of intergalactic deuterium from the dark ages (the focus of their study) because of the reduced backgrounds, but, in followup, \cite{2006PhR...433..181F} asserted that detecting it would require a heroic effort.  

This is the first study to consider the $\Hethreeplus$ hyperfine line in practical detail, and we find that this signal is more detectable than previously thought.  This paper discusses the nature of the intergalactic $8.7~$GHz signal, the sensitivity requirements of an observation to detect it, and this line's potential scientific applications. We find that neither spin-exchange collisions with electrons nor the scattering of \HeII\ Lyman-series photons in the $z\sim 3$ IGM are sufficient to decouple the $\Hethreeplus$ \Hethreesplitting\ spin temperature significantly from the CMB temperature.   This is in contrast to the $21$~cm transition of hydrogen, where calculations find that the spin temperature is coupled to the kinetic temperature by the ambient radiation for $z \lesssim 15$ \citep{2006PhR...433..181F}.  Because of the small spin temperature of intergalactic $\Hethreeplus$, it is easiest to observe this transition in absorption against radio-bright, high-redshift quasars.  

For $2 < z < 6$, the $\Hethreeplus$ hyperfine line falls between $1.2$~GHz and $2.9$~GHz.  This is a higher frequency range than that of redshifted $21$~cm surveys, and is convenient for several reasons.  First, the sky temperature (which represents the noise floor) at $\sim 2\;$GHz is $4\;$K \cite{1995ApL&C..32....7B}, which is much smaller (by a factor $>100$) than that of $z>6$ $21\,$cm emission.  Frequencies of a couple GHz are also less susceptible to ionospheric fluctuations or galactic Faraday rotation than measurements at hundreds of MHz.   Further, these frequencies are not high enough that the lowest major resonances from the atmosphere or from astrophysical sources interfere.  We show that constraints on the $^3{\rm He}$ abundance and \HeI\ reionization are formally within reach of a $10^5\;$m$^2$ interferometer, while a statistical detection of $8.7\,$GHz $\Hethreeplus$ absorption from during \HeII\ reionization could be feasible with existing instruments.

This paper calculates the strength of the $\Hethreeplus$ hyperfine absorption signal (Sec.~\ref{sec:signal}) and discusses its observability.  Sec.~\ref{sec:detection} discusses the sensitivity requirements to detect this signal in absorption.  Finally, Sec.~\ref{sec:contaminants} argues that other redshifted lines which fall into this band are likely to be subdominant to the $\Hethreeplus$ \Hethreesplitting\ absorption and shows that many relevant instrumental considerations are not as stringent as those to detect high-redshift $21~$cm emission.

\section{The $\Hethreeplus$ signal}
\label{sec:signal}

\subsection{The optical depth}
\label{ss:opticaldepth}

The $I=1/2$ nuclear spin of $\Hethreeplus$ splits the ground state into an $F=0$ singlet and an $F=1$ triplet, but because the nuclear moment has the opposite sign relative to hydrogen, the singlet is the excited state \cite{2006JETP..102..367K}.  The optical depth to redshift through the hyperfine line is ~\cite{2004ApJ...608..622Z}  
\beq
\tau_{\Hethreeplus} = \frac{c^3 \hbar A_{10} }{16 \kB \Tspin \nu_{10}^2} \frac{n_{^3{\rm He}^+}}{(1+z) {dv/dy}}, 
\label{eqn:tau}
\eeq 
where $\nu_{10} =$ \Hethreesplitting, $n_{^3{\rm He}^+} = \nH f_{^3{\rm He}} \xHeII \Delta_b$, $\Delta_b$ is the gas density in units of the cosmic mean, $v$ is the proper line-of-sight velocity (Hubble flow plus peculiar velocity), and $y$ is the conformal distance.  The spin temperature $\Tspin$ determines the relative abundance of the hyperfine excited state to the ground state as $\exp[-h \nu_{10}/(\kB \,\Tspin)]/3$, and $\xHeII$ is the fraction of helium that is \HeII\ (and $\bar{x}_{\rm HeII}$ is its volume average).  Lastly, $f_{^3{\rm He}}$ is the fractional abundance of \Hethree\ relative to hydrogen, which we take to be $1.04 \times 10^{-5}$.  This is the value for a $\Lambda$CDM cosmology with $\Omega_b=0.046$, $\Omega_m= 0.27$, zero spatial curvature, massless neutrinos, a Hubble parameter of $h=0.71$, and CMB temperature of $\Tcmb = 2.73\, (1+z)$~K \citep{2009ApJS..180..330K}; we adopt this cosmological model throughout.
   
For a reference epoch, we take $z=3.6$ ($1.84$~GHz, or $16.3$~cm) where the fiducial model for \HeII\ reionization in \citep{2009ApJ...694..842M} predicts $\bar{x}_{\rm HeII} \sim 0.5$, and this prediction derives mainly from the observed quasar luminosity function.  With these choices, we can rewrite Eq.~(\ref{eqn:tau}) as
\beqa
\tau_{\Hethreeplus} &=& 4.5\times 10^{-7} \xHeII \Delta_b \frac{\Tcmb(z)}{\Tspin} \left ( \frac{1+z}{4.6} \right )^{1/2} \nonumber \\
&\times& \left(\frac{H(z)/(1+z)}{dv/dy} \right).
\label{eq:opticaldepth}
\eeqa
Note that in emission this transition has brightness temperature $(\Tspin - \Tcmb) \,\tau_{\Hethreeplus}$, which Sec.~\ref{ss:spintemp} demonstrates is negligible for intergalactic $\Hethreeplus$.  It is most promising to look for the intergalactic signature in absorption against bright continuum sources.

Systems that have overdensities greater than several have decoupled from the Hubble flow, and one can picture these regions as a forest of discrete clouds.  A cloud with column density $N_{\Hethreeplus}$ in $\Hethreeplus$ has
\beq
\tau_{\Hethreeplus} = 1.1 \times 10^{-5} \, \left(\frac{N_{\Hethreeplus}}{10^{15} \, {\rm cm^{-2}}} \frac{10 \, {\rm K}}{\Tspin}\, \frac{30 \, {\rm km \, s}^{-1}}{\Delta v} \right),
\label{eqn:tau2}
\eeq
where $\Delta v$ is the velocity width of the cloud (including Doppler broadening).
From the \HI\ Ly$\alpha$ forest, we measure the neutral hydrogen column density $N_{\rm HI}$ of these clouds, so it is more intuitive to write Eq.~(\ref{eqn:tau2}) in terms of $N_{\rm HI}$, rather than $N_{\Hethreeplus}$.  If we assume that all the helium is \HeII, photoionization equilibrium for the hydrogen, and the analytic formula that maps $N_{\rm HI}$ to $\Delta_b$ given in \citet{schaye01}  (valid for $N_{\rm HI} \lesssim 10^{18}\,$cm$^{-2}$  and which provides an excellent fit to numerical simulations), then the optical depth to \Hethreesplitting\ absorption of a single structure can be written as
\beqa
\tau_{\Hethreeplus} &\approx& 3 \times 10^{-6} \,\left(\frac{N_{\rm HI}}{10^{15} \, {\rm cm^{-2}}} \, \frac{\Gamma_{\rm HI}}{10^{-12} \, {\rm s}^{-1}}\right)^{1/3} \\ \nonumber  & \times & \left( \frac{10 \, {\rm K}}{\Tspin} \, \frac{30 \, {\rm km \, s}^{-1}}{\Delta v} \right) \left(\frac{T}{2 \times 10^4 \, {\rm K}}\right)^{0.5},
\label{eqn:tauofNHI}
\eeqa
where $\Gamma_{\rm HI}$ is the \HI\ photoionization rate, which is measured to be $\Gamma_{\rm HI} \approx 10^{-12} \, {\rm s}^{-1}$ at $z = 2-5$ (e.g. \citep{faucher08}).  Column densities of $10^{15}\,$cm$^{-2}$ are commonplace in the \HI\ Ly$\alpha$ forest ($\sim 1$ per $10$ cMpc at $z\approx 3$). 

\subsection{The spin temperature}
\label{ss:spintemp}

The brightness temperature of $\Hethreeplus$ emission scales as $(\Tspin - \Tcmb)/\Tspin$, while the optical depth to absorption scales as $1/\Tspin$.  Therefore, whether the resonance is viewed in absorption or emission depends on how decoupled $\Tspin$ is from $\Tcmb$.  The spin temperature of the hyperfine transition is primarily determined by: 1) radiative excitation at $\nu_{10}$, 2) collisional spin exchange, 3) excitation by the radiation near \HeII\ Ly$\alpha$ and other allowed transitions (through the $\Hethreeplus$ equivalent of the Wouthuysen-Field effect) \cite{1959ApJ...129..536F}\footnote{The principle of the Wouthuysen-Field coupling is that, when photons scatter off of the $\Hethreeplus$ resonances, each event has some probability to switch the hyperfine state of the ion.  If sufficient scatterings take place, this scattering can couple the $\Tspin$ to the color temperature of the radiation.  In addition, energy exchange between the gas and \HeII\ Lyman-series photons tends to drive the color temperature around the resonance frequency to the gas temperature.}.  Appendix~\ref{apx:additionalspin} summarizes additional, subdominant spin coupling mechanisms.

In equilibrium, the spin temperature is given by 
\beq
\Tspin^{-1} = \frac{\Tcmb^{-1} + x_{\rm c} \Tkinetic^{-1} + x_\alpha T_\alpha^{-1}}{1+ x_{\rm c} + x_\alpha},
\label{eqn:Ts}
\eeq
where $\Tkinetic$ is the kinetic temperature, $T_\alpha$ is the effective color temperature of the resonant radiation, and the $x$'s are coupling coefficients that are given below.  Because of the large difference between $\Tkinetic$ and $T_\alpha$ versus $\Tcmb$, for relevant values of the $x$'s Eq.~(\ref{eqn:Ts}) can be rewritten as
\beq
\Tspin \approx \Tcmb \, (1+ x_{\rm c} + x_\alpha).
\label{eq:Tsapprox}
\eeq

Eq.~(\ref{eqn:Ts}) and Eq.~(\ref{eq:Tsapprox}) assume the CMB is the dominant background at $\nu_{10}$.
\citet{2009arXiv0901.0559S} found mild distortions to the CMB temperature from extragalactic emission at radio frequencies, but this emission is negligible at \Hethreesplitting\ in the IGM at all redshifts~\footnote{Higher density regions could have synchrotron or free-free emission at \Hethreesplitting\ in excess of $\Tcmb$, but these regions likely already have $\Tspin \gg \Tkinetic$ anyway due to collisional coupling.}.

\subsubsection{Collisional spin exchange}
\label{sss:collisional}

Spin exchange from collisions with electrons dominates over spin exchange from collisions with ionic and atomic species, which have thermal velocities that are further suppressed relative to electrons by their masses.  The collisional spin-exchange cross section derived from partial wave phase shifts from \cite{1982PhRvA..26.1401K} is 
\beqa
\sigma(E) &=& \frac{3 \pi}{4 k^2(E)} \sum_{\ell=0}^\infty (2\ell+1) \sin^2[\delta^t_\ell(E) - \delta^s_\ell(E)] \nonumber \\ 
&\approx& \frac{1.05}{k^2(E)}, %= 0.35 \cdot \frac{E_{\rm Ry}}{E} a_o^2, C falls as $\sim 1\%~{\rm eV}^{-1}$
\label{eqn:scatteringsigmaE}
\eeqa
where $\hbar k = \sqrt{2 m_e E}$ is the wavenumber of the scattering electron and we have worked in the approximation that the phase shifts are constant over relevant energies (see Appendix~\ref{apx:secollisions} for additional discussion regarding Eq.~\ref{eqn:scatteringsigmaE}).  The parameter $C$ in $\sigma(E) = C/k^2(E)$ has been evaluated at $1$~eV, and it rises gradually to $1.08$ at $8$~eV. %, and then falls to $\sim 0.9$ by $30$~eV.  
The s and p-wave shifts ($\ell= 0$ and $1$) dominate the summation in Eq.~(\ref{eqn:scatteringsigmaE}).  In addition, the thermally averaged cross section is
\beq
\bar \sigma = \frac{1}{(\kB \Tkinetic)^2} \int_0^\infty dE \sigma(E) E e^{-E/(\kB \Tkinetic)} \approx \frac{14.3~{\rm eV}}{\kB \Tkinetic} a_o^2,
\label{eqn:electronspinexrate}
\eeq
where $a_o$ is the Bohr radius. The collisional coupling coefficient is then \cite{2007MNRAS.374..547F}
\beqa
x_{\rm c} &=& \frac{\nelectron T_\star}{A_{10} \Tcmb} \sqrt{\frac{8 \kB \Tkinetic}{\pi m_e c^2}} c \, \bar \sigma \\ \nonumber 
&\approx& 1.0 \times 10^{-2} \, \Delta_b \, \left(\frac{1 +z}{4.6} \right)^2 \, \left(\frac{\Tkinetic}{10^4 \, {\rm K}} \right)^{-1/2},
\eeqa
where $\nelectron$ is the electron density and the second line assumes that hydrogen is fully ionized and the helium is singly ionized.  The temperature of the IGM is measured from the \HI\ Ly$\alpha$ forest to be $\sim 2\times10^4$~K near mean density at $z\sim 2-4$ \cite{2000ApJ...534...41R, 2000MNRAS.318..817S}. Therefore, the collisional spin coupling will be weak at $z = 3.6$ except in overdense regions ($\Delta_b \gtrsim 100$ or $n_e \gtrsim 3\times 10^{-3}$~cm$^{-3}$ for $x_{\rm c} > 1$). 

\subsubsection{Radiative spin coupling}
\label{sss:radcoupling}

In addition to collisions, ultraviolet radiation produced by  \HeIII$\rightarrow$\HeII\  recombinations and directly by quasars can penetrate into the \HeII\ regions and affect $\Tspin$ by pumping the hyperfine states through electronic dipole transitions.  This effect turns out to be weaker for the $8.7\;$GHz transition of $\Hethreeplus$ compared to the analogous mechanism for the $21$~cm line of atomic hydrogen because $A_{10}$ is much larger in the case of $^3$He+ and because the radiation backgrounds are much weaker at the $\Hethreeplus$ allowed dipole transitions. 

Photons that have energies greater than \HeII\ Ly$\alpha$ can redshift into the \HeII\ Ly$\alpha$ resonance or higher series resonances. However, scattering off of the higher series resonances is not an important contribution to pumping $\Hethreeplus$ hyperfine states because such photons are removed by $^4{\rm He}^+$ before they redshift into the $\Hethreeplus$ resonances \cite{2006ApJ...651....1C}:  A \HeII\ Ly$n$ photon scatters on average $5$ times by  $^4 {\rm He}^+$ before being destroyed, and the optical depth for a photon that redshifts across the $^4$He+ Ly$\alpha$ resonance is $\tau_{\rm ^4HeII, Ly\alpha} = 4200 \, \xHeII \, \Delta_b \, [(1+z)/4.6]^{3/2}$ ($\tau_{\rm ^3HeII, Ly\alpha} = 0.5 \, \xHeII \, \Delta_b \, [(1+z)/4.6]^{3/2}$ for the $^3$He+ Ly$\alpha$ resonance).    Approximately $0.3$ of each destroyed Ly$n$ photon for $n >2$ ends up as a \HeII\ Ly$\alpha$ photon, which contributes to the pumping ~\citep{2006MNRAS.367..259H, 2006MNRAS.367.1057P}.  
%Photo-electric absorption of these photons by HI or by metals is typically negligible.

Coupling of the spin temperature to the color temperature of the radiation at $\Hethreeplus$ Ly$\alpha$ is described by the coefficient \cite{field58, 2004ApJ...602....1C, 2006MNRAS.367..259H}
\beq
x_\alpha = \frac{P_{10} T_\star}{A_{10} \Tcmb} = \frac{16 \pi \chi_{\rm HeII}^\alpha T_\star}{9 A_{10} \Tcmb} \beta_\infty S_\alpha = \frac{\beta_\infty}{\beta_\alpha} S_\alpha,
\eeq
where
 \beq
\beta_\alpha = 9.6 \times 10^{-10} \left ( \frac{1+z}{4.6} \right ) \; {\rm cm}^{-2} \, {\rm Hz}^{-1} {\rm s}^{-1} {\rm sr}^{-1},
\eeq
$P_{10} = (4/9) \, P_{\alpha, \Hethreeplus}$ is the scattering de-excitation rate \footnote{This can be derived from Eq. (23) in \citet{field58} noting that the level structure is the same as hydrogen, except that the singlet energies are above the triplet, which means that $P_{10}/P_{\alpha, \Hethreeplus}$ for $\Hethreeplus$ is the same as $P_{01}/P_{\alpha, H}$ for hydrogen, where $P_{01}$ is the respective scattering excitation rate.}, $P_{\alpha, \Hethreeplus} \equiv 4 \pi \chi^{\alpha}_{\rm HeII} \beta_\infty S_\alpha$ is the \HeII\ Ly$\alpha$ scattering rate, $\chi_{\rm HeII}^\alpha = \pi e^2 /(m_e c) f_\alpha$ is the \HeII\ Ly$\alpha$ cross section with oscillator strength $f_\alpha =  0.416$, and $\kB T_\star \equiv h \nu_{10}$.  The function $S_\alpha \equiv \int d\nu \phi(\nu) \beta(\nu)/\beta_{\infty}$ (where $\phi(\nu)$ is the absorption profile normalized such that $\int d\nu \phi(\nu) = 1$) describes the shape of the spectrum near resonance.  The spectrum is not changed significantly around resonance at relevant $T_{\rm K}$ such that $S_\alpha \approx 1$.

Ly$\alpha$ scattering off of $^4$He+ rather than $^3$He+ shapes the spectrum of the radiation at the $\Hethreeplus$ Ly$\alpha$ resonance, which is offset $14$~km s$^{-1}$ redward of the $^4$He+ Ly$\alpha$ resonance or a few $\nu_D$ for $10^4$\,K gas, where $\nu_D$ is a Doppler width of $^4$He+.  In the absence of redshift, repeated scattering off of $^4$He+ creates a ``mesa'' absorption profile in the incident spectrum centered on $^4$He+ Ly$\alpha$ with half-width $(2 \log \tau_{\rm ^4HeII, Ly\alpha})^{1/2} \, \nu_D \sim 4 \, \nu_D$ \citep{1959ApJ...129..536F}.  Redshifting helps to extend the mesa profile across the $^3$He+ Ly$\alpha$ resonance.  Energy exchange with the gas during scattering causes the slope of the flat part of the mesa to converge to that of the Planck function with temperature equal to $\Tkinetic$  \citep{1959ApJ...129..536F}, but whether $T_\alpha \approx \Tkinetic$ at the $^3$He+ resonance depends on the temperature, ionization state, and density of the gas.  Furthermore, absorption of ultraviolet photons by the $3$d~$^3$P$^{0}_{2}$--$2$P$^2$~$^3$P$_2$ resonance of {O{\sc ~iii}} that falls $9$~km s$^{-1}$ redward of $^3$He+ Ly$\alpha$ can also deplete the flux redward of this resonance \citep{1985ApJ...290..578D}.  This effect can drive $T_s$ to negative values \cite{1985ApJ...290..578D}.  Although, this effect has only been considered for oxygen abundances that are orders of magnitude larger than the mean IGM abundance.   Because of these considerations, a detailed treatment is required to determine $T_\alpha$.  Fortunately, such a treatment is not required for this study because we find $x_\alpha \ll 1$.  However, even if $x_\alpha$ were near unity, the value of $T_\alpha$ would not be important for determining $T_s$ except in locations where $T_\alpha$ is fine-tuned such that $|T_\alpha| \lesssim \Tcmb$, which is unlikely.

%We find that as long as $|T_\alpha| \gg \Tcmb$, equation (\ref{eq:Tsapprox}) will be valid and therefore $T_s$ does not depend on $T_\alpha$ and is only affected by ultraviolet photons through $S_\alpha$ and $\beta_\infty$.  Since it is unlikely that $|T_\alpha| \sim \Tcmb$ and these processes do not significantly affect $S_\alpha$ such that $S_\alpha \approx 1$, the most relevant parameter to characterize ultraviolet pumping is $\beta_\infty$, which we estimate below.

Scatterings from \HeII\ Ly$\alpha$ photons produced by recombinations pump the \Hethreesplitting\ transition. However, a  \HeIII$\rightarrow$\HeII\  recombination must take place $\sim .1~$cMpc from the edge of the \HeII\ region in order for any \HeII\ Ly$\alpha$ photon that results to be able to reach the \HeII\ region edge before redshifting off resonance \footnote{Scattering in the wing of the Ly$\alpha$ line is significantly less important for $z \sim 3$ helium than $z > 6$ hydrogen because of its lower abundance, the lower densities at these redshifts, and a photon redshifts $4$ times faster across the \HeII\ Ly$\alpha$ line.}.  Therefore, only if the \HeII\ region contains some \HeIII\ gas (or if the recombining gas is near the \HeII\ region) are \HeII\ Ly$\alpha$ photons produced by recombinations present.  It is likely that a large fraction of the ionizing photons during \HeII\ reionization have long mean free paths and can penetrate deep into \HeII\ regions, producing \HeIII~\citep{1998ApJ...501...15M, 2009ApJ...694..842M}. 

An estimate for the specific intensity (in units of ${\rm cm}^{-2} {\rm s}^{-1} {\rm Hz}^{-1} {\rm sr}^{-1}$) at \HeII\ Ly$\alpha$ from {\it in situ} recombinations is 
\beqa
\beta_\infty \biggl |_{\rm rec} &\approx& 0.7 \, \frac{c \, \alpha_B(T) \bar{n}_e \bar n_{\rm He}}{4 \pi} \langle \Delta_b^2 \xHeIII \rangle \,  \left |\frac{dt}{dz} \frac{dz}{d\nu} \right | \label{eqn:beta}\\ \nonumber 
&\approx& 6 \times 10^{-13} \left ( \frac{1+z}{4.6} \right )^{9/2} \left(\frac{T}{10^4 {\rm  K}} \right)^{-0.7} \langle \Delta_b^2 \xHeIII \rangle,
\eeqa
where $\alpha_B$ is the Case B \HeII\ recombination coefficient, and $0.7$ represents the $\approx 0.7$ Ly$\alpha$ photons, on average, that each \HeIII$\rightarrow$\HeII\ recombination produces~\cite{1989agna.book.....O}.  The brackets signify the average over the diffusion scale [this scale must be smaller than the total distance a photon travels as it redshifts across resonance or $0.1 (\Delta v/10 \, {\rm  km \, s^{-1}})$ cMpc and larger than $\tau_{\rm ^4HeII, Ly\alpha}^{-1/2}$ times this distance].  The value of $\langle \Delta_b^2 \xHeIII \rangle$ in most intergalactic \HeII\ regions is less than unity because $\Delta_b \sim 1$ and $\xHeIII \ll 1$.  In the simulations of \citet{2009ApJ...694..842M}, at the time that $50\%$ of the helium in the IGM is \HeII, the mean values of $\xHeIII$ in \HeII\ regions is $\sim 0.1$ (and becomes $\sim 0.3$ when there is net $80\%$ ionization of \HeII).  Assuming $x_{\rm HeIII} = 0.5$, the required value of $\Delta_b$ such that $x_\alpha \gtrsim 1$ is $\Delta_b \gtrsim 40$ at $z = 3.6$.  Therefore, recombinations are not sufficient except in very overdense regions to decouple $T_{\rm S}$ from $T_{\rm CMB}$.

Higher series Lyman photons can also pump this transition.  However, only a small fraction $n>2$ \HeII\ Lyman series photons are converted into Ly$\alpha$ photons, and their contribution to $\beta_\infty$ is much smaller than direct \HeII\ Ly$\alpha$ production from recombinations (see \footnote{In a \HeII\ region \HeII\ Ly$n$ photons are either locally scattered by $^4$He+ and destroyed, which $\approx 0.3$ of the time for $n>2$ ($0$ of the time for $n=1$) results in a Ly$\alpha$ photon that can pump the $\Hethreeplus$ (this fraction is included in the $0.7$ in Eq. \ref{eqn:beta}), or it will redshift until it reaches the next resonance before it has a chance of being converted to a \HeII\ Ly$\alpha$ photon.  The contribution to pumping from photons produced by recombinations in \HeIII\ regions would essentially be given by equation (\ref{eqn:beta}), but with the replacement $\langle \Delta_b^2 \xHeIII \rangle \rightarrow 0.02 \,\bar{\Delta_b^2} \;\bar{x}_{\rm HeIII}$, where $0.02$ factor comes from the small probability of producing such photons in a recombination cascade and, then, the probability that these photons produce a \HeII\ Ly$\alpha$ photon.}).

Ultraviolet photons with energies greater than $40.8\,$eV and less than $54\,$eV  that are produced by quasars can also pump the $\Hethreeplus$ hyperfine states since these photons will redshift into the HeII Lyman resonances.  (Stars in galaxies produce very few photons with energies greater than $40.8\,$eV.)  Because quasar spectra should not vary significantly over this wavelength interval, the radiation that redshifts onto resonances is a fairly homogeneous background.  (Note that a photon redshifting from \HeII\ Ly$\beta$ to \HeII\ Ly$\alpha$ travels $600 [(1+z)/4.5]^{1/2}$ cMpc.)  However, detailed estimates for the UV background from quasars in \cite{2009arXiv0901.4554F} find $\beta \lesssim 10^{-12} \, {\rm s^{-1} cm^{-2} \,Hz^{-1} \,  sr^{-1}}$ in this band, which is insufficient to significantly decouple $T_s$ from $T_{\rm CMB}$ \footnote{In addition, several of the $^3$P$^{0}$--$2$P$^2$~$^3$P resonances of O{\sc ~iii} are offset $\sim 100$~km s$^{-1}$ blueward of the $^3$He+ Ly$\alpha$ resonance.  These resonances can destroy continuum photons~\cite{1985ApJ...291..492D}, suppressing $\beta_\infty$ even further.}.\\

In summary, we find that $\Tspin \sim \Tcmb$ except in regions dense enough for collisional coupling to be important ($\Delta_b \gtrsim 100$ at $z = 3.6$ for $x_c = 1$) or in dense regions where radiative coupling can operate ($\Delta_b \gtrsim 40$ at $z = 3.6$  for $x_\alpha = 1$).    In contrast, galactic detections of $\Hethreeplus$ from \HII\ regions have found the $8.7\;$GHz line of $\Hethreeplus$ in emission.  There, though, the electron number density is $\sim 10^8$ times higher than in the $z \sim 3$ IGM.

Because the $\Tspin$ does not decouple significantly from $\Tcmb$, detecting intergalactic $\Hethreeplus$ \Hethreesplitting\ emission during this era would require a mammoth effort.  Even if one were to take $\Tspin \gg \Tcmb$ for all the $\Hethreeplus$ gas in the universe, we find that sensitivities of $\sim 0.3~\mu{\rm K}$ at $\sim10$~cMpc scales would be required.  However, because we showed that $\Tspin \approx \Tcmb$ except in and near galaxies, the sensitivity requirements are at least a factor of several more stringent.  Simple estimates suggest that this small signal still could be detected with the Square Kilometer Array.  However, it is not clear that $\Hethreeplus$ is the dominant emission line at $\sim 2$~GHz and, for example, $4.2$~GHz hyperfine emission from $^{14}$NV may be more important (Sec. \ref{ss:hyperfinemetals}), and $21$~cm emission will dominate over the $\Hethreeplus$ emission signal from $z > 5.1$.  Detecting the $\Hethreeplus$ $8.7$~GHz line in absorption is the focus for the remainder of this paper.

\section{Prospects for detection}
\label{sec:detection}

\begin{figure}
\rotatebox{-90}{\epsfig{file=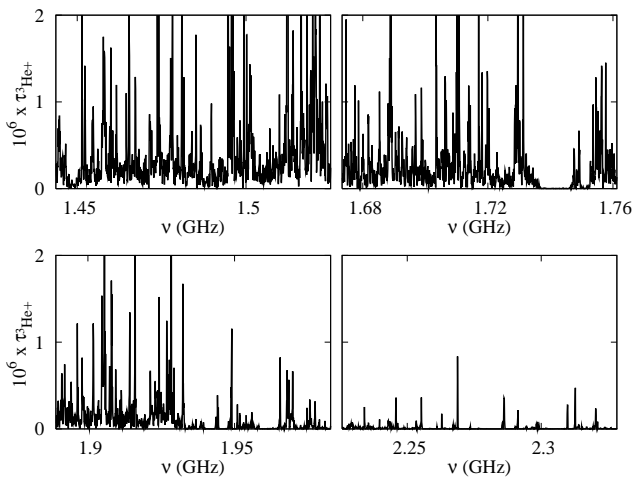, height=8.5cm}}
\caption{The $\Hethreeplus$ absorption signal along four $190$ cMpc skewers at different times during \HeII\ reionization, calculated from simulation L1 in \cite{2009ApJ...694..842M}.  The top-left panel is from $\bar{z} = 5.0$ and volume-averaged \HeII\ fraction of $\bar{x}_{\rm HeII, V} \approx 0.9$, the top-right is from $\bar{z} = 4.2$ and $\bar{x}_{\rm HeII, V} \approx 0.7$,  the bottom-left is from $\bar{z} = 3.6$ and $\bar{x}_{\rm HeII, V} \approx 0.4$, and the bottom-right is $\bar{z} = 2.9$ and $\bar{x}_{\rm HeII, V} \approx 0$. Note that these simulations do not resolve the densest peaks, where $\tau_{\Hethreeplus}$ is largest.\label{fig:skewers}}
\end{figure}

Fig.~\ref{fig:skewers} shows an estimate for the $\Hethreeplus$ absorption signal along four $190$ cMpc skewers and at four times during \HeII\ reionization.  Each skewer is calculated from the L1 simulation in \cite{2009ApJ...694..842M}, assuming that $\Tspin = \Tcmb$.  \HeIII\ regions can be identified from these skewers (the regions in which most pixels have zero flux).  Note that these calculations do not capture the densest, most absorbed regions during \HeII\ reionization, which may remain self-shielded after \HeII\ reionization.  After \HeII\ reionization, estimates are that there will be $\sim 1$ self-shielding system per $30$~cMpc (e.g., \cite{2000ApJ...530....1M}).   In addition, the ionizing radiation from the observed quasar will decrease the amount of absorption close to it, a ``$\Hethreeplus$ proximity effect''.  This effect is not included in  Fig.~\ref{fig:skewers} and may prove valuable for probing the \HeII\ reionization process.

In this section, we first discuss the high-redshift, radio-bright source population (Sec.~\ref{ss:sources}) and then quantify the prospects for detecting the \Hethreesplitting\ absorption signal with existing and planned radio telescopes (Sec.~\ref{ss:sensitivity}).  We discuss three methods to detect the signal: (1) direct imaging of $8.7$~GHz absorption, (2) a statistical detection of it, and (3) a targeted observation that uses the quasar's Ly$\alpha$ forest absorption to locate regions with large $\tau_{\Hethreeplus}$.  We quantify the telescope specifications required to detect the signal with each of these methods.  This section discusses only the formal sensitivity limit from thermal radiometer noise, and Sec.~\ref{ss:instrumental} describes important instrumental factors.

\subsection{The high-redshift, radio-bright quasar population}
\label{ss:sources}

Radio-bright quasars represent $\sim 10\%$ of the quasar population, and their abundance peaks at $z\sim2.5$ \cite{2002A&A...391..509H, 2003ApJ...591...43V, 1997MNRAS.284...85D}.  The radio quasars described here are also optically luminous, allowing a determination of their redshift \cite{2002A&A...391..509H}.  We use the fact they are optically luminous in Sec.~\ref{sss:guidedlya} to consider the \HI\ Ly$\alpha$ signal in parallel.  Many radio observations have identified and studied these rare objects \cite{2003ApJ...591...43V, 2006MNRAS.370.1034C, 2002A&A...391..509H, 1999ApJ...518L..61V, 2007AJ....133.2841C}.  Taken together, surveys for radio-bright quasars cover a significant fraction of the sky, but are heterogeneous in flux limits, completeness, area, and spectroscopic followup in the optical.  It is therefore challenging to estimate space densities of this population.

We found $17$ sources with $z>2$ and $1.4$~GHz fluxes $\gtrsim 200$~mJy from recent radio-bright quasar catalogs \cite{2003ApJ...591...43V, 2006MNRAS.370.1034C, 2002A&A...391..509H, 2004A&A...424...91E}.  Our selection is not complete, and is only meant to indicate that such sources exist.  In the $z \sim 3.5$ range, the luminous ``GHz peaked"~\cite{2004A&A...424...91E} source J1445+0958 at $z=3.53$ with $\sim 2$~Jy at $1.4$~GHz is very promising (one also has PMN J1230-1139 at $z = 3.53$ with $480$~mJy at $1.4$~GHz  and PMN J2003-3252 at $z= 3.78$ with $470$~mJy at $1.4$~GHz).  At lower redshifts, one finds sources such as J0240-2309 at $z=2.23$ with $\sim 5$~Jy at $1.4$~GHz and several other $z\sim2.3$ sources with $\sim 2$~Jy at $1.4$~GHz~\cite{2004A&A...424...91E}.  At the high redshift end, the $z=5.11$ quasar TN J0924-2201 has a flux of $70$~mJy at $1.4$~GHz \cite{1999ApJ...518L..61V, 2007AJ....133.2841C}. 

\subsection{Sensitivity}
\label{ss:sensitivity}

\subsubsection{Directly resolving the $\Hethreeplus$}
\label{ss:imaging}

The RMS thermal noise for an interferometric measurement of the flux of a point source is
 \begin{eqnarray}
\Delta S &\approx & \frac{ 2 \kB \, \gamma \, T_{\rm sys}}{A_{\rm eff} \, \sqrt{2 \, t_{\rm obs} \, \Delta \nu}} \nonumber \\
&=& 0.5~{\rm \mu Jy} \,\cdot \gamma  \left(\frac{T_{\rm sys}}{20\, K} \, \frac{10^5 {\rm m^2}}{A_{\rm eff}} \right) \left( \frac{{\rm week}}{t_{\rm obs}} \frac{\rm MHz}{\Delta \nu} \right)^{\frac{1}{2}},
\label{eqn:radiometer}
\end{eqnarray}
where $A_{\rm eff}$ is the effective collecting area of the interferometer, $\gamma$ encodes the instrumental efficiency, $\Delta \nu$ is the width of the frequency channel, $t_{\rm obs}$ is the integration time, and $T_{\rm sys}$ is the system temperature of the instrument.  Instruments that can currently operate at $2\;$GHz achieve system temperatures of $20$--$30\,$K, which is set by receiver noise rather than the sky ($T_{\rm sky} \approx 4\;$K).  

To directly resolve this absorption in mean density gas at $z = 3.6$ with a week-long observation, an ideal instrument with $A_{\rm eff} = 10^4 \, {\rm m}^2$ requires a source with $S_{\rm 1.8 GHz} = \Delta S/\tau_0  \approx  10$~Jy (roughly five times the flux of any known $z\sim 4$ source), where $\tau_0$ is the $\Hethreeplus$ optical depth for mean density gas in the Hubble flow.  Here we have considered $10^4 \, {\rm m}^2$ to be representative of existing telescopes~\footnote{From their respective proposer's guides and overviews: for the GBT, we take $2$~K/Jy; for the VLA, we take $0.098$~K/Jy; for the GMRT, we take $0.22$~K/Jy.  All gains here are for a single antenna.}, where in the bands we consider: 1) the GBT has $A_{\rm eff} \sim 5500  \, {\rm m}^2$, 2) the GMRT ($\nu < 1.45~{\rm GHz}$) has $A_{\rm eff} \sim 18200  \, {\rm m}^2$ and, 3) the VLA has $A_{\rm eff} \sim 7300  \, {\rm m}^2$.
For contrast, the Square Kilometer Array (SKA) would require a source with only $S_{\rm 1.8 GHz}  \approx 100~$mJy to image this signal.  Overdensity (and infall) can boost the absorption and reduce these requirements.  

Sensitivities of $\Delta S \approx 1\,\mu$Jy in $1\,$MHz spectral bins have not been achieved at these frequencies.  For example, \citet{2002A&A...381L..29O} performed a $5\,$hr observation on the Very Large Array at $1.7$~GHz in search of OH absorption along a sightline, and they achieved $\approx 100\,\mu$Jy RMS per MHz, a factor of $100$ off the mark for directly resolving the $\Hethreeplus$ signal.  However, we show in subsequent sections that clever methods can relax the sensitivity requirements to detect this absorption to $\Delta S \approx 30 \, \mu \,$Jy in spectral bins of $0.1\,$MHz.  

\subsubsection{Detecting the power spectrum of $8.7\;$GHz absorption}
\label{sec:statistical}

The power spectrum of $\tau_{\Hethreeplus}$ provides the simplest statistical description of the $8.7\,$GHz absorption.  To calculate the sensitivity of an instrument to the power spectrum, we need to estimate the power spectrum of the instrumental noise.
The power spectrum of the noise can be derived by equating $k_{\rm N} \, P(k)/ \pi$ with $\Delta S^2$, and is equal to
\beq
P_N(k) = \Delta S^2 \Delta \nu \frac{dy}{d\nu},\label{eqn:PN}
\eeq
where $k$ is the wavevector in comoving units and $k_{\rm N}$ is the Nyquist wavevector, and where, for redshifts at which $\Omega_m(z) \approx 1$,
\beq
\frac{dy}{d\nu}  \approx 2 \left ( \frac{z+1}{4.5} \right )^{1/2}~{\rm cMpc} \cdot {\rm MHz}^{-1}.
\eeq 
If $P_N(k)< P_\tau(k)$, where $P_\tau$ is the power spectrum of $\tau_{\Hethreeplus}(\nu)$, one can image individual $k$-modes.

The signal-to-noise (S/N) ratio of a statistical detection of $P_\tau$ is given by
\beq
\left ( \frac{S}{N} \right )^2 = \frac{1}{2} \,\sum_{\rm modes} \left ( \frac{P_\tau(k)}{P_\tau(k) + P_N(k)/S^2} \right )^2.  
\label{eqn:sn}
\eeq
Eq.~(\ref{eqn:sn}) assumes Gaussianity.  The noise should be Gaussian, but the signal will not be.  When the signal is comparable to the noise, Eq.~(\ref{eqn:sn}) is an overestimate.

The S/N scales with experimental parameters (in the limit that radiometer noise dominates over sample variance) as 
\beq
\frac{S}{N} \propto \frac{B^{1/2} \, A_{\rm eff}^2 \langle S^2 \rangle \, N_{\rm QSO}^{1/2} \, t_{\rm obs}}{\gamma^2 \, \Delta \nu^{1/2} \, T_{\rm sys}^2},
\label{eqn:scalings}
\eeq
where $N_{\rm QSO}$ is the number of quasars in which this absorption is observed, $B$ is the bandwidth, and Eq.~(\ref{eqn:scalings}) assumes that $P_\tau$ is independent of $k$ (which is approximately true for $k \lesssim 10$~cMpc$^{-1}$).

\begin{figure}
\rotatebox{-90}{\epsfig{file=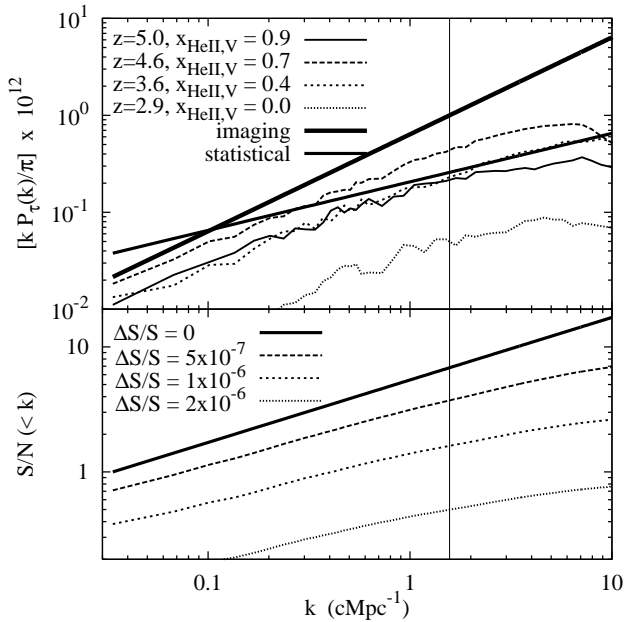, height=8.4cm}}
\caption{Top panel: The power spectrum of $\tau_{\Hethreeplus}$ at several times during \HeII\ reionization, estimated from the L1 simulation of \HeII\ reionization presented in \cite{2009ApJ...694..842M}. Also included is the noise power spectrum (the curve labeled ``imaging''), assuming an observation with $\Delta S/S = 10^{-6}$ at $1\,$MHz over $B = 100\,$MHz.  If the noise power is smaller than $P_{\tau_{\Hethreeplus}}$, individual modes in the signal can be imaged.  The ``statistical'' curve is the error on $P_\tau$ in a bin of size $\Delta k/k = 0.3$. The vertical line corresponds to the Nyquist wavevector for $1\,$MHz spectral resolution.  Bottom Panel:  The cumulative S/N ratio as a function of wavevector for the $z=3.6$ signal shown in the top panel, assuming $B = 100\,$MHz and the specified $\Delta S/S$ in $1\,$MHz spectral bins. \label{fig:power}}
\end{figure}

The top panel in Fig.~\ref{fig:power} shows $P_\tau$ at $4$ times during the L1 simulation of \HeII\ reionization presented in \cite{2009ApJ...694..842M} under the assumption $\Tspin = \Tcmb$ (which is an excellent approximation; Sec.~\ref{ss:spintemp}).  It is clear that the amplitude changes during \HeII\ reionization, but the shape does not evolve considerably, even though the \HeIII\ bubbles grow substantially over the redshift range considered.  Aliasing makes the growth of bubbles less apparent in the 1-D power spectrum (and other statistics may be better suited to discriminate the bubble growth and structure).  The ``imaging'' curve in Fig.~\ref{fig:power}  is the sensitivity of an observation with $\Delta S/S = 10^{-6}$ at $1$ MHz (Eq. \ref{eqn:PN}), and the ``statistical' curve is the error on $P_\tau$ in bins of $\Delta k/k = 0.3$ for an observation with $B = 100$ MHz (including the cosmic variance term using the signal from $z\approx 3.6$).   

The bottom panel in Fig.~\ref{fig:power} calculates the S/N ratio (including cosmic variance) for different sensitivities in $1\,$MHz spectral bins assuming $B = 100\,$MHz and using the $z=3.6$ signal shown in the top panel.  This panel demonstrates that an observation with $\Delta S/S = 10^{-6}$ in  $1\,$MHz pixels is required to detect $P_\tau$ at $S/N \sim 3$ and that an observation that is twice as sensitive yields most of the available information (compare the $\Delta S/S = 5\times 10^{-7}$ curve with the $\Delta S/S = 0$).

\subsubsection{A guided observation that uses the Ly$\alpha$ forest}
\label{sss:guidedlya}

The sensitivity requirements to detect intergalactic \Hethreesplitting\ absorption are relaxed if the Ly$\alpha$ forest absorption (as well as Ly$\beta$ and Ly$\gamma$) can also be measured for the same intergalactic systems.  Lyman-series absorption reveals where the densest regions in the IGM lie, such that the signal can be measured just from the pixels in which the S/N ratio is expected to be highest.  This strategy can result in a large gain in an observation's sensitivity to \Hethreesplitting\ absorption.  In Sec.~\ref{sec:statistical}, we saw that an observation with $\Delta S/S \sim 10^{-6}$ for $\Delta \nu = 1$ MHz was needed to detect this signal.  Here we show that $\Delta S/S \approx 3\times 10^{-5}$ for $\Delta \nu = 0.1$ MHz in combination with Ly$\alpha$ forest data can yield a significant detection of gas that will be ionized by \HeII\ reionization, and in Sec. \ref{sec:largeNHI} we show that dense systems could provide an even larger signal.

Using the \HI\ Ly$\alpha$ forest as a template, an optimistic estimate for the S/N ratio that an observation of $8.7$~GHz absorption can achieve is
\beq
\left(\frac{S}{N}\right)_{\rm opt}^2 = \bar{x}_{\rm HeII}^2 \, \Delta z \int _{N_{\text{HI}}^{\min}}^{N_{\text{HI}}^{\max }}\frac{\text{d$^2$ N}}{\text{dz} N_{\rm HI}} \left(\frac{S \, \tau_{\Delta \nu}}{\text{$\Delta
$S}}\right)^2 d N_{\rm HI}, 
\label{eqn:optimistic}
\eeq
where ${\text{d$^2$N}}/{\text{dz} N_{\rm HI}} \approx 3\times 10^7 \, N_{\rm HI}^{-1.5} \, (1+z)^{2.46}$ is the column-density distribution of $N_{\rm HI}$ using the fit in \citep{press93} valid for $z> 1.7$ and $N_{\rm HI} < 2 \times 10^{17}$~cm$^{-2}$,  $N_{\text{HI}}^{\min }$ and $N_{\text{HI}}^{\max}$ are the minimum and maximum \HI\ column densities included in the measurement, and $\Delta z$ is the redshift interval that the observation spans.   In addition, $\tau_{\Delta \nu} \approx \sigma_{\Hethreeplus} N_{^3 {\rm He}}/\Delta \nu$, where $\sigma_{\Hethreeplus} = c^2 h A_{10}/(32 \pi \nu_{10} k_B T_{\rm CMB})$, is the effective optical depth to the \Hethreesplitting\ hyperfine transition if the absorber falls within a spectral pixel of width $\Delta \nu$ and has $x_{\rm HeII}= 1$.   Finally, we assume $\Gamma_{\rm HI} = 10^{-12}$ s$^{-1}$ \citep{faucher08} and we use the same mapping between $N_{\text{HI}}$ and $N_{\Hethreeplus}$ used to derive Eq.~(\ref{eqn:tauofNHI}), which yields the scaling $N_{\Hethreeplus} \propto N_{\text{HI}}^{1/3}$.  

Equation (\ref{eqn:optimistic}) is an estimate for the average S/N ratio that an observation of \Hethreesplitting\ absorption can achieve, assuming that the location of all absorbers with $N_{\rm HI}$ between $N_{\rm HI}^{\rm max}$ and  $N_{\rm HI}^{\rm min}$ is known (via the Lyman forest).  In practice, the \HI\ Ly$\alpha$ forest saturates easily (and the other Lyman resonances are contaminated by absorption from Ly$\alpha$ absorption at lower redshift and span a smaller redshift interval) such that it is difficult to discriminate between absorbers with $N_{\rm HI}$ in the range $10^{14}$--$10^{19}$cm$^{-2}$.  

A more realistic estimate for the S/N ratio may be to assume that an observation co-adds the radio signal from the location of dense absorbers with $N_{\rm HI} > N_{\rm HI}^{\min}$ (i.e., does not use knowledge about the value of $N_{\rm HI} $).  In this case, an estimate for the S/N ratio at which $\Hethreeplus$ $8.7$~GHz absorption can be detected is
\beqa
\left(\frac{S}{N}\right)_{\rm cons}^2 &=& \frac{\bar{x}_{\rm HeII}^2 \, \Delta z}{\Delta S^2} \, \left(\int _{N_{\text{HI}}^{\min}}^{N_{\text{HI}}^{\max }}\frac{\text{d$^2$N}}{\text{dz} N_{\rm HI}} \,( S \, \tau_{\Delta \nu}) \, d N_{\rm HI} \right)^2 \nonumber \\
&\times&  \left(\int _{N_{\text{HI}}^{\min }}^{N_{\text{HI}}^{\max }}\frac{\text{d$^2$N}}{\text{dz} N_{\rm HI}} \, d N_{\rm HI} \right)^{-1}. \label{eqn:cons}
\eeqa
Therefore, equation (\ref{eqn:optimistic}) represents an $N_{\rm HI}$-weighted estimate for the S/N ratio and equation (\ref{eqn:cons}) represents an unweighted estimate.

Ly$\alpha$ forest absorbers with $N_{\rm HI} > 10^{13}$ cm$^{-2}$ have widths of order $10-60$~km~s$^{-1}$ \citep{press93, theuns98}.  Note that $\Delta \nu = 100$ kHz  at $2\,$GHz corresponds to $15$ km s$^{-1}$.  Lines with $N_{\rm HI} > 10^{15-16}\,$cm$^{-2}$ will self-shield even after \HeII\ reionization \citep{2009ApJ...694..842M} and have optical depths of unity to \HeII\ Lyman-limit photons.  Systems that are moderately self-shielded are still highly ionized, such that it is likely safe to consider absorbers with $N_{\text{HI}}^{\max} \lesssim 10^{16}$ cm$^{-2}$ as probing \HeII\ reionization.  Also, note that $\Delta_b \approx 150 \, (N_{\rm HI}/10^{17} {\rm~cm}^{-3})^{2/3}$ in the assumed model \cite{schaye01}, such that $x_c \approx 1$ for $N_{\rm HI} = 10^{17} {\rm~cm}^{-3}$.  Therefore, electron collisions increase $T_s$ by a factor of $2$ over $T_{\rm cmb}$ at this column density and by a larger factor above it, reducing $\tau_{\Hethreeplus}$.

\begin{figure}
{\epsfig{file=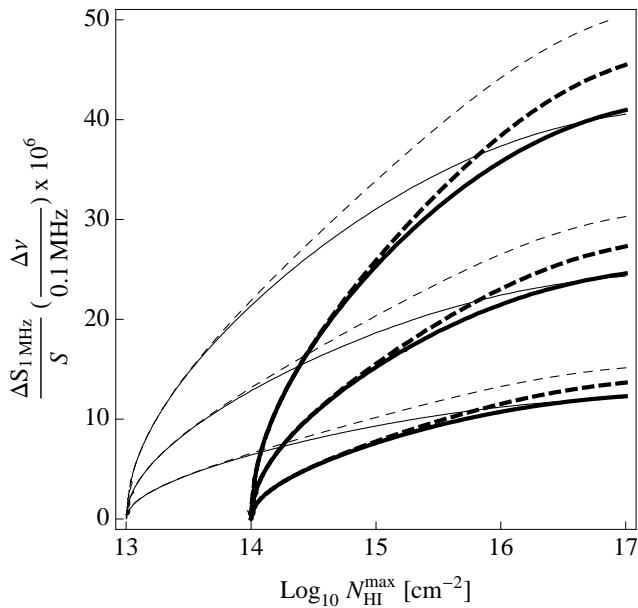, width=8.4cm}}
\caption{Contours for  $3$, $5$ and $10$ $\sigma$ detections in the $N_{\rm HI}^{\rm max}$  versus $[\Delta S/S \; (\Delta \nu/ 0.1 \, {\rm MHz})]$  plane.  The dashed curves are for the optimistic estimator and the solid are the conservative one.  We assume for this calculation $z=3.6$, $N_{\rm HI}^{\rm min} = 10^{13}$ cm$^{-2}$ (thin curves) or $N_{\rm HI}^{\rm min} = 10^{14}$ cm$^{-2}$ (thick curves), that $\bar{x}_{\rm HeII} = 1$, and $B = 100\,$MHz. \label{fig:stat_lya}}
\end{figure}

Fig.~\ref{fig:stat_lya} shows our optimistic (dashed curves; Eq. \ref{eqn:optimistic}) and conservative (solid curves; Eq. \ref{eqn:cons}) estimates for the significance that $\Hethreeplus$ \Hethreesplitting\ absorption could be detected with this method if $\bar{x}_{\rm HeII} = 1$.  We assume for this calculation $z=3.6$, $B= 100\;$MHz ($\Delta z \approx 0.25$), and $N_{\rm HI}^{\min} = 10^{13}$ cm$^{-2}$  (thin curves) or $N_{\rm HI, min} = 10^{14}$ cm$^{-2}$  (thick curves).  Both the optimistic and conservative estimates yield similar results, namely that $\Delta S \sim 30\, \bar{x}_{\rm HeII}$~$\mu{\rm Jy}$ is required to detect the \Hethreesplitting\  absorption signal using absorbers with $N_{\rm HI} \lesssim 10^{16}\,$cm$^{-2}$ for $S = 1~$Jy.   Also, note that it becomes easier to use the \HI\ Ly$\alpha$ forest absorption to find dense regions as the redshift decreases (and the forest becomes more transparent), and it would be difficult to use this absorption to guide $8.7$~GHz absorption studies at $z \gtrsim 5$ where only underdense regions are unsaturated.

Fig.~\ref{fig:stat_lya} shows that observations with $\Delta S/S \gtrsim 2\times10^{-5}$ are most sensitive to the highest column density lines ($N_{\rm HI} \gtrsim 10^{16}\,$cm$^{-2}$).   However, for such observations there will be significant sightline-to-sightline scatter in the significance of a detection:  There are $\sim 1$ systems with $N_{\rm HI} > 10^{17}$cm$^{-2}$ per $100$~MHz and this number scales as  $N_{\rm HI}^{-0.5}$ with $N_{\rm HI}$.

\subsubsection{$\Hethreeplus$ absorption in super Lyman-limit and damped Ly$\alpha$ systems}
\label{sec:largeNHI}

Thus far we have focused mostly on $\Hethreeplus$ absorption in intergalactic absorption systems with $N_{\rm HI} \lesssim 10^{16}\,$cm$^{-2}$.  Higher column-density systems are not as affected by \HeII\ reionization.   However, clouds with $N_{\rm HI} \gtrsim 10^{20}$ cm$^{-2}$ (called ``DLAs'') are easily identifiable in the \HI\ Ly$\alpha$ forest because the damping wings of the hydrogen line appear in absorption.  There will be $\sim 1$ such system per unit redshift at $z \approx 3$ \citep{press93}.

These systems may make the easiest targets to search for large values of $\tau_{\Hethreeplus}$, and, since they will appear in absorption even after \HeII\ reionization, programs that aim to find redshifted $\Hethreeplus$ absorption may find the most immediate success if they target low-redshift DLAs, utilizing the brightest extragalactic radio sources on the sky.   

It is thought that between $N_{\rm HI} \sim 10^{17}\,$cm$^{-2}$ and $N_{\rm HI} \sim 10^{21}\,$cm$^{-2}$, the nature of these absorbers shifts from being overdense intergalactic regions to being systems that live within halos and in galactic disks.  Unfortunately, systems with $N_{\rm HI} \gtrsim 10^{18}\,$cm$^{-2}$ are not as well-understood as those with lower $N_{\rm HI}$ because hydrodynamic simulations of the IGM do not include the detailed radiative transfer needed to model the ionization state of dense, self-shielding systems.  It is not obvious how much \HeII\ absorption is expected from these high column-density systems and such modeling is beyond the scope of this paper.    It is possible that some dense systems have $\tau_{\Hethreeplus} \sim 10^{-4}$ (especially at low redshift where $\Tcmb$ is smaller).  See footnote \footnote{For DLAs, most of the hydrogen should be self-shielded to ionizing photons such that the hydrogen gas is substantially neutral.   Because of this, the ratio of $N_{\rm HI}$ to $N_{\Hethreeplus}$ will be smaller than for lower $N_{\rm HI}$ systems.  However, the helium in \HI\ regions will often stays in the \HeII\ state because the photoionization cross section for \HeI\ is a much flatter function of frequency than that of the \HI.  Thus, hard photons penetrate into \HI\ regions and keep the helium singly ionized.  An effect that suppresses the $\Hethreeplus$ absorption signal is that at particle densities of $n =3\times 10^{-3}\,$cm$^{-3}$ in \HII\ regions (and of $n =4\times 10^{-2}\,$cm$^{-3}$ in \HI\ regions) $\Tspin = 2 \, \Tcmb$.  Henceforth, $\Tspin$ increases approximately linearly with density such that $\tau_{\Hethreeplus}$ becomes independent of density.} for further discussion.

\subsection{$\Hethreeplus$ absorption from $z > 5.1$}
\label{sec:reion}
Lower redshift $21$~cm emission contaminates the $\Hethreeplus$ signal at $z  > 5.1$.  A mere $10^8\, M_\odot$ in neutral hydrogen  with $20$ km s$^{-1}$ circular velocity at $z = 0.3$ has $S \approx 5 \,\mu$Jy.  It is conceivable that this large contaminant could prohibit the use of $\Hethreeplus$ \Hethreesplitting\ absorption at high redshift.  The most obvious application for \Hethreesplitting\ absorption is to study \HeI\ reionization (which should be concurrent with hydrogen reionization).  

Fortunately, low-redshift $21$~cm emission is confined to galaxies, and, therefore, a small enough interferometric beam may be able to peer through the holes in the galaxy distribution and detect this high-redshift \Hethreesplitting\ absorption signal.  The $z = 0$ \HI\ luminosity function constructed with the HIPASS survey extends to galaxies that contain $10^7\, M_\odot$ in neutral hydrogen, with a number density of detected objects of order $0.1$ cMpc$^{-3}$ \citep{2003AJ....125.2842Z}.  If we conservatively assume a galaxy number density of $1$ cMpc$^{-3}$, to observe \Hethreesplitting\ absorption at $z = 7$ (or $21$~cm radiation from $z=0.3$) requires a beam constructed with baselines of length $\sim 1$ km in order for $1$ foreground galaxy on average to enter the beam per $10\,$MHz.  In addition, if we assume that all galaxies have radius $10$~kpc in this toy model, the ultimate limit that a long baseline observation could achieve is $\sim 0.01$ foreground $21$~cm-emitting galaxy per $10\,$MHz.  

Observations of reionization of the first electron of helium via $\Hethreeplus$ \Hethreesplitting\ absorption would require a bright $z >6$ source.  It is possible that a bright enough population of sources exist.  \citet{2007AJ....133.2841C} observed two $z =5.2$ radio galaxies that had fluxes $74\,$mJy and  $18\,$mJy at $1.4\,$GHz to search for high-redshift $21$~cm absorption.  These sources were also selected to have a soft spectral index, so it is possible that even brighter $z>5$ sources exist at $\sim 2\,$GHz.   Under ideal conditions, a $1$ week observation of a $50\,$mJy source with the SKA would be sensitive to $1~\mu {\rm Jy}$ absorption features.

\section{Contaminants}
\label{sec:contaminants}

\subsection{Instrumental sources of contamination}
\label{ss:instrumental}

There are a few significant challenges to the proposed program of wideband precision spectrometry: 1) terrestrial radio frequency interference (RFI), 2) passband calibration and stability, 3) gain calibration across polarizations, 4) distortions owing to a turbulent ionosphere, and 5) the frequency dependence of the instrumental beam.  While it is beyond the scope of this study to quantify the severity of these challenges, we argue that many of these challenges are less severe for observations of \Hethreesplitting\ absorption compared to $z>6$ $21$~cm emission.

The relevant spectral band has many broadband transmissions and satellite downlinks, with only small regions formally limited to radio astronomy~\footnote{The FCC only formally allocates $1.66$~GHz to $1.67$~GHz to radio astronomy and experimental downlinks.  For a survey of RFI at the GBT see L-band and S-band RFI studies under {\tt http://www.gb.nrao.edu/IPG/}.}.  While geographic avoidance of sight lines is helpful~\cite{2006NewA...11..551F}, RFI presents significant challenges to a wideband measurement.  Transmissions will also be a concern for redshifted $21$~cm observations.  Because observations of $\Hethreeplus$ absorption will only target a small patch of the sky and can rely on the correlations between the longest baselines whereas the sensitivity of redshifted $21$~cm observations is dominated by short baselines of $\lesssim 100~$m \citep{2006ApJ...653..815M}, it may be easier to control this systematic for \Hethreesplitting\ absorption.  

Passband calibration is another significant consideration.  While the error from thermal noise decreases with the square root of the integration time, variations in the passband (caused by instrumental resonances, or variations in the spectrometer passband) represent an irreducible noise term that could remain after usual calibration techniques such as chopping on and off-source.  Because the signal power spectrum spans a wide range of spatial wavevectors, statistical detection may benefit slightly if passband errors are restricted to particular frequencies or wavevectors.

A third systematic, gain differences between the two linear polarizations, could lead to spurious signal.  This systematic is more severe for cosmological $21$~cm observations.  For an observation of $21$~cm emission at $200$ MHz over a bandwidth of $10$ MHz, the polarization vector of the foregrounds will rotate $2~$rad for rotation measure (RM) equal to $10^{-3} \, {\rm rad\, cm^{-2}}$ (a rough value for RM above the galactic plane \cite{2004mim..proc...13J}) whereas for \Hethreesplitting\ emission at $2$ GHz over $100$ MHz it will rotate $0.02~$rad.  Therefore, such gain miscalibration will lead to more line-of-sight structure for redshifted $21$~cm emission.

Fourth, the ionosphere will distort the signal on timescales that are as small as seconds.   However, the ionospheric distortion is much smaller for redshifted $8.7~$GHz absorption than that of redshifted $21$~cm~\cite{2004WRM....14S.189X}.  Deflections through the ionosphere are less than $4''$, compared to $0.16^\circ$ at $250$~MHz.  

Finally, the beam of any instrument is a function of frequency, such that additional sources enter in the beam as the frequency decreases.  If the beam is sufficiently smooth with frequency, any leakage from sources entering and exiting the beam will contribute a smooth component on top of the signal that can be subtracted off.  Since the severity of this leakage enters as a $\Delta \nu/ \nu$ effect, this systematic is also likely to be less of a concern for $\Hethreeplus$ observations than for that of redshifted $21\,$cm:  Redshifted $21$~cm measurements are most sensitive to $\gtrsim 10~$cMpc modes for which $\Delta \nu/ \nu \lesssim 100$ \cite{2006ApJ...653..815M} whereas $\Hethreeplus$ observations are sensitive to $\lesssim 1~$Mpc modes  for which $\Delta \nu/ \nu \gtrsim 1000$.

\subsection{Line emission contaminants}
\label{ss:linecontam}

Since the abundance of $\Hethreeplus$ is not much larger than the universal abundance of certain metals, contamination from different lines is a considerable concern.  Our galaxy is the strongest source for many of the possible contaminating lines.
However, terrestrial and galactic transitions can in principle be masked in searches for the redshifted $\Hethreeplus$ absorption \footnote{
The frequencies under consideration fall below the most notable molecular resonances in the atmosphere, and are well above the plasma frequency of the ionosphere, so these are both mild considerations compared to lower and higher-frequency experiments.  In this frequency range, the atmosphere's effective temperature is $\sim 1$~K \cite{1995ApL&C..32....7B, 1989ApJ...342..604D} and is dominated by emission from molecular oxygen (which is very flat in frequency \cite{1989ApJ...342..604D}), but depends on precipitable water vapor at the level of $1$~mK/mm.  Further, in the bulk of the atmosphere, pressure broadening wipes out spectral structure on the scales considered.
}.  Extragalactic lines are weaker, but cosmological redshift allows higher frequency transitions to fall into the desired range and smears their effect over a range in frequency.  In what follows we argue that the most likely contaminants are subdominant to the $\Hethreeplus$  signal.

\subsubsection{Hyperfine transitions from metals}
\label{ss:hyperfinemetals}

For hydrogenic hyperfine transitions, the wavelength depends on the inverse cube of $Z$~\cite{2007AstL...33...67S, 1984SvAL...10..201S}. Therefore, hydrogen-like ($1s$) transitions are at much higher frequencies, and only lithium-like ($2s$), boron-like ($2p_{1/2}$), and sodium-like ($3s$) ionic states fall at frequencies that can contaminate the \Hethreesplitting\ signal.  There are a few effects that suppress these contaminants: (1) isotopes that have nuclei with non-zero spin are generally rare, (2) the abundances of metals is low (especially in the IGM at high redshift where the average metallicity is $\sim 10^{-3}$ Solar), and (3) generally only a fraction of the metals are in the ionization state that could contaminate the \Hethreesplitting\ signal.

We find that the most threatening hyperfine lines are the $4.2$ GHz transition of $^{14}$NV ($A = 9.8\times 10^{-14} \, s^{-1}$, with isotopic fraction of 99.6\%, and mass fraction $1.3 \times 10^{-3} \, W \, \Omega_b$, where $W$ is the abundance relative to Solar), the $5.6$ GHz transition of  $^{13}$CIV ($A = 1.7\times 10^{-13} \,s^{-1}$, 1.1\%, $7.3 \times 10^{-5} \, W \, \Omega_b$), and the $5.0$ GHz transition of $^{29}$SiV ($A = 6.3\times 10^{-13} \, s^{-1}$, 4.7\%, $4.4 \times 10^{-5} \, W \, \Omega_b$).  Of these transitions, $^{14}$NV will be biggest concern.  However, it should be a subdominant absorber to $^3$He in the IGM because its Einstein-A is $20$ times smaller and its number density is only comparable for $W= 0.1 \gg \langle W_{\rm IGM} \rangle$.  Also, the spin temperature is probably coupled to the gas temperature for $^{14}$NV (as well as for other non-hydrogen-like hyperfine transitions), suppressing its absorption (see \footnote{Collisions that change the valence electron state between the nS and nP electronic states in non-hydrogen-like ions may efficiently pump these hyperfine lines at IGM densities \citep{1984SvAL...10..201S}, resulting in emission.  Therefore, redshifted $^{14}$NV may provide an important contribution to the diffuse background at $\sim 2$ GHz at the $\delta T_b \sim 10^{-7}$~K-level, assuming $\langle W \rangle = 0.1$.}).  

\subsubsection{Radio recombination lines}
\label{ss:radiorecomb}

Radiation emitted and absorbed through transitions in extremely excited states of hydrogen -- radio recombination lines (RRLs) -- has been considered as a possible contaminant for \HI\ $21$~cm radiation (e.g., \citep{2003MNRAS.346..871O}).  The RRLs of \HI\ fall at frequencies
\beq
\nu \approx 2 \; \GHz \; \Delta n \; \left(\frac{n}{150}\right)^{1/3}
\eeq
where $n$ is the principal quantum number.  If a line falls within a frequency channel of width $\Delta \nu$ it contributes an optical depth \citep{1975A&A....43..465S}
\beq
\tau_{\rm RRL} = 2\times10^{-10} \, \left(\frac{T_{\rm k}}{10^4 \, {\rm K}}\right)^{-\frac{5}{2}} \left( \frac{ {\rm EM}}{1 {\rm cm^{-6} \, pc}} \right) \left(\frac{\Delta \nu}{1 \, \MHz}\right)^{-1},
\eeq
where we have assumed local thermodynamic equilibrium (LTE) and EM is the emission measure.  Note that it can be dangerous to assume LTE for RRLs.  Collisions drive the occupation number of the highest $n$ states closest to LTE, but lower $n$ states will often be depopulated relative to LTE.  This inversion can lead to enhanced emission or absorption from the highest $n$ states than in LTE (and even maser emission).  However, \citet{1975A&A....43..465S} showed that the corrections from LTE are most important at lower frequencies than relevant here.

H$\alpha$ surveys find that in the direction of the galactic poles ${\rm EM} \sim 1 \, {\rm cm^{-6} \, pc}$ \citep{2003ApJS..146..407F}, and the contribution to the EM from extragalactic objects will be comparable.  Studies of damped \HI\ Ly$\alpha$ systems in the Ly$\alpha$ forest reveal that a typical sightline passes through the outskirts of a few galaxies \citep{2005ARA&A..43..861W}, and so we expect $\langle {\rm EM} \rangle \sim 10 \, {\rm cm^{-6} \, pc}$.  Therefore,  $\tau_{\rm RRL}$ should be much less than $\tau_{\Hethreeplus}$.  Note that a sightline may pass through a galactic \HII\ region where $\tau_{\rm RRL}$ is much larger.  However, because the typical size for the radio emitting region of a quasar is $\sim 1\,$kpc, \HII\ regions in an external galaxy can cover only a small fraction of the quasar radio beam. 

RRL emission should also be unimportant.  The radio recombination line emission from a $10\,$kpc disk at a distance of $1000$ cMpc contributes $10^{-4}\; \mu {\rm Jy}$ when averaged over a $1$ MHz pixel, assuming ${\rm EM} = 1 \, {\rm cm^{-6} \, pc}$ for all sightlines that intersect the disk and $T_{\rm k} = 10^4\,$K.

\subsubsection{Molecular Lines}
\label{ss:molecules}

Rotational transition rates of molecules are generally many orders of magnitude higher than for $\Hethreeplus$.   Fortunately, the lowest lying rotation lines of diatomic molecules rotation lines lie at $\gtrsim 50$ GHz, and would enter our $1-2$ GHz  band at redshifts at which their abundance is negligible.  However, the redshifted signal of hyperfine lines from OH at $\sim 1.7$~GHz \cite{1977A&A....58..403K} or rotational modes of polyatomic molecules like H$_2$CO at 4.83~GHz, CH$_3$OH at 6.67~GHz and 12.18~GHz, ${\rm H}_2{\rm O}$ at 22.2~GHz \cite{1957IAUS....4...92T}, and ${\rm NH}_3$ at 23.7~GHz fall into our band (although, the latter two transitions would originate from $z>10$).  

The abundance of polyatomic molecules even in molecular clouds is many orders of magnitudes lower than $^3$He (with the exception of H$_2$O), and we find that their absorption signal is subdominant to $\Hethreeplus$ except where the sightline directly passes through a galaxy.  At these intersections, OH absorption at rest frame frequencies of $1.5-1.7$ GHz is likely the biggest concern (contaminating the $8.7~$GHz absorption for $z >4$).  Molecular emission lines from galaxies that enter the beam can also contaminate the signal, but we estimate that their signal is again unimportant if LTE holds.

However, these low frequency molecular lines are often not found to be in LTE and several such lines are associated with maser emission ($T_{\rm line} < 0$) and can be extremely bright. In particular, OH $^2 \Pi_{3/2}$ state hyperfine transitions in galaxies from $z=0-0.2$ are a possible contaminant of the $\Hethreeplus$  signal at $1.5-1.7$ GHz ($\Hethreeplus$ absorption at $4.1 <  z <5.1$).  These masers are concentrated in massive galaxies that are ultra-luminous in the infrared (because far infrared radiation pumps these masers).   A small enough interferometric beam avoids significant contamination from these rare galaxies, and the beam size requirements are much less stringent than to avoid low redshift $21$~cm emission outlined in Sec.~\ref{sec:reion}.  

The other potentially important known molecular maser transition is the $6.67$~GHz transitions of H$_2$CO.  However, the only identified extragalactic H$_2$CO masers are in the Large Magellanic Cloud (with $L \approx 10^{-5} - 10^{-6} \, L_\odot$), and despite H$_2$CO maser searches toward M33 and strong OH maser galaxies, only upper limits have been placed on their luminosities \citep{2003AJ....125.1177D}.

\subsubsection{Fine structure lines}
\label{ss:finestructure}
Absorption by fine structure lines can occur from an electronic state with $n>1$ provided it has no downward allowed $\Delta n \ge 1$  transitions, which would depopulate this state.  We concentrate only on hydrogen because of its abundance.  For hydrogen, only the 2S state meets the above criteria
through the $2P_{3/2} \rightarrow 2S_{1/2}$ transition at $2.7$ cm  ($10.9$ GHz), which has spontaneous transition rate $A_{\rm FS} = 8.9 \times 10^{-7} \; {\rm s}^{-1}$ \citep{1952ApJ...115..206W}.  This transition can appear in absorption in \HII\ regions (if the 2S state is pumped by recombinations or collisions) or in \HI\ regions (if it is pumped with \HI\ Lyman-$\beta$ absorption followed by H$\alpha$ emission).  However, \citet{2008MNRAS.390.1430D} showed that the later
mechanism is extremely inefficient.  Furthermore, in an ionized IGM, the Gunn-Peterson optical depth for this transition is $\tau_{\rm FS} \sim  10^{-15} \,  [(1 +z)/{4} ]^{9/2} \Delta_b^2$.
Therefore, intergalactic $2P_{3/2} \rightarrow 2S_{1/2}$ absorption by hydrogen is completely negligible compared to that of \Hethreesplitting\ $\Hethreeplus$.

In galactic \HII\ regions, collisions and recombinations mediated by the higher densities will more efficiently pump the 2S state, and it is possible that $2P_{3/2} \rightarrow 2S_{1/2}$ absorption will exceed our signal.  Along a sightline such regions will be rare and will be associated with DLAs in the \HI\ Ly$\alpha$ forest, allowing these absorbers to be masked.

\section{Conclusions}
\label{sec:conclusion}

We have considered the $^2 {\rm S}_{1/2}$ $F=0-1$ \Hethreesplitting\ hyperfine transition of $^3{\rm He}^+$ as an observable of the ionization state, density structure, and composition of the intergalactic medium.  Despite the $\sim 10^{-5}$ deficit in abundance of $^3{\rm He}$ relative to hydrogen, this transition is compelling because of 1) the high spontaneous rate, which exceeds the $21$~cm rate by a factor of 680, 2) the convenient frequency range over which the redshifted signal falls, which has much lower sky temperatures compared to the frequencies for $z \gtrsim 6$ $21\,$cm emission, and 3) the techniques needed to extract this signal are in sharp contrast with that of redshifted $21$~cm emission.

We showed that collisional and radiative processes are insufficient in the IGM at relevant redshifts to couple $\Hethreeplus$ to $\Tkinetic$ or $T_\alpha$, and, thus, $\Tspin \approx \Tcmb$.  Since $\Tspin$ does not decouple significantly from $\Tcmb$ at relevant densities ($T_s \lesssim 2 \, \Tcmb$ for $\Delta_b < 100$ at $z = 3.6$), detecting intergalactic $\Hethreeplus$ \Hethreesplitting\ emission from this era would require a mammoth effort, but the $\Hethreeplus$ \Hethreesplitting\ line may be observable in absorption along the sightline to bright quasars with present and upcoming instruments.

An obvious science application of $\Hethreeplus$ hyperfine absorption is to study \HeII\ reionization.  This process is believed to occur at lower redshift than hydrogen reionization, with several lines of evidence indicating that it completes by $z\sim3$.  However, all present observations that claim a detection of \HeII\ reionization are controversial because of their indirect nature.  The most direct present-day probe of this process is the \HeII\ Ly$\alpha$ forest, but this absorption saturates at \HeII\ fractions of $10^{-3}$ at the cosmic mean density.  

In contrast, intergalactic hyperfine absorption of $^3{\rm He}^+$  is a linear tracer of the product of the density and the fraction of helium that is \HeII.  This absorption against the brightest sources at $z \sim 4$ (which have $S \sim 1\,$Jy) creates $\sim 1 \;\mu$Jy fluctuations on $\gtrsim 0.1\;$MHz scales.  While such sensitivities are beyond the reach of present-day interferometers, we showed that a $10^5$~m$^2$ instrument operating at its thermal limit could directly resolve these fluctuations, and the SKA could potentially image these fluctuations with high signal to noise. 

A key insight that reduces the sensitivity requirements for detection is that \HI\ Ly$\alpha$ forest absorption at optical wavelengths can be used to locate the spectral bins with large values of $\tau_{\Hethreeplus}$ (in overdensities).  Using this information to stack pixels in a radio observation, one can statistically detect $8.7$~GHz absorption with an RMS noise of $\sim 30 \, \mu$Jy in $0.1\,$MHz spectral bands over $100$ MHz ($\Delta z \approx 0.25$) on a $1~$Jy source.  These sensitivity requirements may be within the reach of present interferometers.

Furthermore, at $z \lesssim 3$ current instruments may be able to study $^3{\rm He}^+$ absorption in systems that are self-shielded to \HeII-ionizing photons against the brightest radio sources on the sky.  Estimates are that there are $\sim 10$ such systems per $100$ MHz, and \HI\ Ly$\alpha$ forest information can also be exploited to stack such systems.   At present, there is considerable uncertainty with modeling $8.7$~GHz absorption from the most dense systems ($N_{\rm HI} \gtrsim 10^{18}$~cm$^{-2}$), and further study is required to understand how much absorption is expected.  However, the densest systems, which will appear as DLAs in the Ly$\alpha$ forest, could have $\tau_{\Hethreeplus} \sim 10^{-4}$.

Another application of intergalactic $8.7$~GHz absorption is to constrain the primordial $^3$He abundance.  To achieve such a constraint, an interferometer would target intergalactic \HeII\ regions along skewers, and from these directly measure the abundance of $^3$He, analogously to how $\Omega_b$ is measured from the \HI\ Ly$\alpha$ forest.  Such observations would need to image the signal ($\sim \mu$Jy sensitivities in $\sim 1\,$MHz spectral bins).  The primordial $^3$He abundance is currently constrained by observations of galactic \HII\ regions, which involves detailed modeling of the \HII\ region lines and is also uncertain to the extent stellar nucleosynthesis effects the abundance in these regions.   An intergalactic measurement of the $^3$He abundance does not suffer from these uncertainties and, therefore, could be more decisive.

Finally, an even more ambitious project would be to use this absorption to study the reionization of the first electron of helium, which is thought to occur at the same time as the reionization of hydrogen.  To study the reionization of the first electron of helium using $8.7$~GHz absorption with the SKA would require a $\sim 50\,$mJy source at $z>6$ to image this signal (assuming perfect instrumental calibration and $T_{\rm sys}= 20\,$K).  Source populations at high-redshift are very uncertain, but we argued that it is possible that such sources exist.  Low-redshift $21\;$cm emission falls in the same band and can contaminate such observations.  However, an interferometric measurement with $\gtrsim 1$~km baselines should be able to avoid significant contamination from this foreground.

Further study of precision wideband spectrometry in this (super-$21\,$cm) range may prove worthwhile. The dominant radiation at these frequencies is the well-characterized CMB such that it may be possible to isolate interesting foreground signals.  Potential observables in this band include other intergalactic hyperfine lines~\cite{1957IAUS....4...92T}, molecular and fine-structure lines from the first galaxies~\cite{basu04, 2008A&A...489..489R}, and spectral distortions produced during cosmological recombination~\cite{2006MNRAS.371.1939R}.\\

\begin{acknowledgments}
M.M. and E.S. contributed equally to the writing and calculations presented in this study.  We would like to thank Chris Hirata and Steve Furlanetto for useful comments on the manuscript.  We acknowledge discussions with Hsiao-Wen Chen, Simon DeDeo, Mark Dijkstra, C.-A. Faucher-Gigu{\`e}re, Lars Hernquist, Chris Hirata, Adam Lidz, Ann Mao, Daniel Marrone, Jeff McMahon, Kenneth Nollett, Brant Robertson, and Matias Zaldarriaga.  E.S. thanks participants of the workshop, ``The Physics of Cosmological Recombination'' for discussion and stimulating study of signals in this frequency range, and is, in particular, grateful to Chris Hirata for suggesting $\Hethreeplus$ in the context of the recombination spectral distortions.  M.M. acknowledges support from the NSF.  E.S. acknowledges support by NSF Physics Frontier Center grant PHY-0114422 to the Kavli Institute of Cosmological Physics.  \\

While this project was nearing completion, we learned of a similar effort on 3He+ hyperfine emission by \citet{bagla09}.  We refer the reader there for complementary discussion.

\end{acknowledgments}

\appendix
\section{Additional spin temperature-coupling processes}
\label{apx:additionalspin}

In addition to collisional and radiative processes, there are four other processes which are subdominant: 1) magnetic dipole interaction with a passing electron, and interactions between the proton spin and a passing electron 2) bound-free and free-bound processes, 3) collisions that drive electronic state transitions or resonant scattering, and 4) charge exchange \cite{1984SvAL...10..201S}.  However, the magnetic dipole interaction between the ionic electron and passing electron as well as the interaction between the proton and passing electron is negligible \cite{2007MNRAS.374..547F}. % also 1985PhRvA..31.2854F
Thermal electron capture at low $Z$ is also subdominant to the thermal spin exchange \cite{1989ZPhyD..14..317A}.

Electronic excitation of the $\Hethreeplus$ from a passing electron can perform a similar role as radiative spin pumping.  To excite $1s \rightarrow 2p$, one needs $E_{\rm th} = (3/4) Z^2 E_{\rm Ry} \approx 40.8~{\rm eV} \gg \kB \Tkinetic$ (where $E_{\rm Ry}=13.6$~eV), which is clearly scarce at relevant temperatures of $\sim 10^4$~K.  Additionally, an electron can scatter resonantly through an intermediate state.  Here, the passing electron becomes temporarily bound, introducing $E_{\rm free}+E_{\rm binding}$ to excite the ground state.  We follow \cite{1984SvAL...10..201S} in estimating the contribution from this process.  First consider the impact excitation for a free electron.  Here, the cross section is relatively flat above threshold $E_{\rm th}$ \cite{1997PhRvA..55..329F}, and 
\beqa
\bar \sigma_{\rm 1s-2p} &=& \frac{1}{(\kB \Tkinetic)^2} \int_{E_{\rm th}}^\infty dE \sigma(E) E e^{-E/(\kB \Tkinetic)} \nonumber \\
&\approx& \frac{9~{\rm eV}}{\kB \Tkinetic} a_o^2 e^{-E{\rm th}/(\kB \Tkinetic)}.
\eeqa
The resonant (intermediate bound state) form of this process is roughly boosted by $\exp(Z^2 E_{\rm Ry}/(4 \kB \Tkinetic)$.  This is still clearly insufficient at IGM temperatures (where $\kB \Tkinetic \approx 1~{\rm eV} \ll (2/3) E_{\rm th})$ to boost the spin exchange rate.

In charge exchange, ${\rm He}^+ + {\rm H} \leftrightarrow  {\rm He} + {\rm H}^+ + \gamma$.  The forward rate gives an upper bound on the spin exchange through this mechanism, and is \cite{1998ApJ...509....1S}
\beq
\frac{\dot x_{\rm HeII}}{x_{\rm HeII}} \approx -(1 \times 10^{-15}~{\rm cm}^{3} s^{-1} ) \left ( \frac{\Tkinetic}{300~{\rm K}} \right )^{1/4} n_H x_{\rm HI},
\eeq
which, during this era, is $ \sim 10^{-24}~{\rm s}^{-1}$, and so is also negligible.  In addition, charge exchange with atoms other than H should be even more negligible.

\section{Electron-$^3{\rm He}+$ scattering}
\label{apx:secollisions}

Here, we consider the spin exchange from electron-$\Hethreeplus$ collisions (described in Sec.~\ref{sss:collisional}) in more detail.  For an incident plane wave, the output wave has an unscattered and a scattered component \cite{2007MNRAS.374..547F},
\beq
\Psi_{\rm out} \propto e^{ikx} + f_k(\theta) \frac{e^{i kr}}{r},
\label{eqn:outstate}
\eeq 
such that the differential cross section is $d \sigma / d \Omega = | f_k(\theta) |^2$.  We index the spin state using bra-ket notation with placeholders nuclear spin, comma, atomic electron spin, scattering electron spin.  In analogy to \cite{field58}, but applied to $\Hethreeplus$, the atomic wavefunction for the singlet excited state is ${\psi_H/\sqrt{2} [ {| \uparrow, \downarrow \cdot \rangle} - {| \downarrow, \uparrow \cdot \rangle} ]}$ where $\psi_H$ is the space part of the wave function.  An unpolarized incident electron has wavefunction ${e^{ikx}/\sqrt{2} [ e^{i \gamma} {| \cdot, \cdot \uparrow \rangle} + {| \cdot, \cdot \downarrow \rangle} ]}$.  The combined spin space of the incident electron and singlet-state atom that will produce a spin exchange is ${ \psi_H e^{ikx}/2 [ e^{i\gamma} | \uparrow, \downarrow \uparrow \rangle - | \downarrow, \uparrow \downarrow \rangle ]}$.  We can write the electron-electron component of the interaction as the sum of triplet and singlet states, ${| \uparrow, \downarrow \uparrow \rangle} = \frac{1}{2} [{(| \uparrow, \downarrow \uparrow \rangle + | \uparrow, \uparrow \downarrow \rangle)} - {(| \uparrow, \uparrow \downarrow \rangle-| \uparrow, \downarrow \uparrow \rangle)]}$.  Through the interaction, though, the electron-electron singlet and triplet will be mapped differently to output states as 
\beqa
{| \uparrow, \downarrow \uparrow \rangle} &\mapsto& \frac{1}{2} [T(\theta) {(| \uparrow, \downarrow \uparrow \rangle + | \uparrow, \uparrow \downarrow \rangle)} \nonumber \\  &-& S(\theta) {(| \uparrow, \uparrow \downarrow \rangle-| \uparrow, \downarrow \uparrow \rangle)]},
\eeqa
and analogously for ${| \downarrow, \uparrow \downarrow \rangle}$.  Here, $T(\theta)$ and $S(\theta)$ describe scattering through angle $\theta$ between the asymptotic in and out states in the momentum basis. If we now find the matrix element to the final state $m_F=1$ of the $F=1$ atomic triplet output state $\langle \uparrow, \uparrow \cdot |$ to get the scattering part of the matrix element
\beq
\frac{e^{ikr} \psi_H}{4 r} [ e^{i \gamma} (T(\theta)-S(\theta)) | \cdot, \cdot \downarrow \rangle ] \Rightarrow \frac{d \sigma}{d \Omega} = \frac{1}{2^4} |T(\theta)-S(\theta)|^2.
\label{eqn:diffxsection}
\eeq
(Note that this calculation simplifies the wave functions to anticipate terms that contribute to the $m_F=1$ or $m_F=-1$ final state cross sections \cite{field58}.  The restriction to spin-flip wavefunctions does not apply to $m_F=0$.)
Expand T and S identically in the partial wave basis as \cite{1962PhRv..126..163B}
\beq
T(\theta) = \frac{1}{2 i k} \sum_\ell (2 \ell + 1) [ e^{2 i \delta^t_\ell}-1] P_\ell (\cos(\theta)).
\eeq
This relates the momentum space representation of $T(\theta)$ and $S(\theta)$ to an angular momentum basis in which the scattering matrix is diagonal, making unitarity manifest through real phase shifts $\delta^s_\ell$ and $\delta^t_\ell$.  Integrating $d \sigma / d \Omega$ over angle and using the orthogonality of the Legendre polynomials gives
\beq
\sigma = \frac{\pi}{4 k^2} \sum_\ell (2 \ell +1) \sin^2(\delta_\ell^t - \delta_\ell^s).
\eeq
Taking the total cross section to any of the triplet state produces the factor of $3$ in Eq.~\ref{eqn:scatteringsigmaE}.  

To evaluate $\sigma$, we use the calculated values for the phase shifts in \cite{1982PhRvA..26.1401K}.   However, even without knowledge of these values, one can derive a unitarity bound assuming that s-wave scattering dominates.  Note that s-wave scattering dominates the spin-exchange cross section at relevant energies \cite{2007MNRAS.374..547F}.  Since $\sin^2(\delta_0^t - \delta_0^s) < 1$, this yields the bound
\beq
\sigma < \frac{3 \pi}{4 k^2} \Rightarrow \bar \sigma <  \frac{3 \pi}{4} \frac{13.6~{\rm eV}}{k_B T} a_o^2.
\eeq
This bound is only $30\%$ larger than our more detailed estimate for $\bar \sigma$ (Eq.~\ref{eqn:electronspinexrate}).  

All of the interaction physics in our calculation of $\sigma$ is hidden in $\delta_l^t$ and $\delta_l^s$.  Employing a more physically complete approach, \citet{1989ZPhyD..14..317A} decomposed the scattering matrix into a direct (Rutherford) Coulomb scattering term and an exchange term.  They then proceeded to evaluate the exchange term using the Coulomb wave function for the scattering electron.  They found that $\sigma \propto Z^{-2}/E$ for hydrogen-like ions, where $Z$ is the nuclear charge.  Thus, the effectiveness of spin exchange decreases with nuclear charge.  

%This is in contrast to capture, which scales as $Z^2$.  Indeed, the relative effectiveness of capture and spin exchange cross over at $Z\sim 25$ \cite{1989ZPhyD..14..317A}. 

\bibliography{heliumthree_hyperfine}
\end{document}